\documentclass[pre,aps,onecolumn,superscriptaddress,floatfix]{revtex4}

\usepackage{graphicx}
\usepackage{dcolumn}
\usepackage{bm}
\usepackage{color}

\usepackage{amsmath}
\usepackage{amssymb}
\usepackage{latexsym}
\usepackage{multirow}
\usepackage{xcolor}
\usepackage{hyperref}
 
\newcommand{\xtil}{\widetilde{x}}

\newcommand{\xitil}{\widetilde{\xi}}

\usepackage{textcomp}

\usepackage{hyperref}
\hypersetup{colorlinks=true,linkcolor=cyan,citecolor=cyan}

\begin{document}

\title{Long-term memory induced correction to Arrhenius law}

\author{A. Barbier-Chebbah$^{1,2}$, O. B\'enichou$^2$, R. Voituriez$^{2,3}$, T. Gu\'erin$^{4}$}
\affiliation{$^{1}$ Decision and Bayesian Computation, USR 3756 (C3BI/DBC) and Neuroscience Department CNRS UMR 3751, Institut Pasteur, Universit\'e de Paris, CNRS, 75015 Paris, France}
\affiliation{$^{2}$Laboratoire de Physique Th\'eorique de la Mati\`ere Condens\'ee, CNRS/UPMC, 4 Place Jussieu, 75005 Paris, France}
\affiliation{$^{3}$Laboratoire Jean Perrin, CNRS/UPMC, 4 Place Jussieu, 75005 Paris, France}
\affiliation{$^{4}$Laboratoire Ondes et Mati\`ere d'Aquitaine, CNRS/University of Bordeaux, F-33400 Talence, France}

\bibliographystyle{naturemag}

\date{\today}

\begin{abstract} 
The Kramers escape problem  is a paradigmatic model for the kinetics of rare events, which are usually characterized by Arrhenius law. So far, analytical approaches have failed to capture the kinetics of rare events in the important case of non-Markovian processes with long-term memory, as occurs in the context of reactions involving proteins, long polymers, or strongly viscoelastic fluids.
Here, based on a minimal model of non-Markovian Gaussian process with long-term memory, we determine quantitatively the mean FPT to a rare configuration and provide its asymptotics in the limit of a large energy barrier $E$.
Our analysis unveils a  correction to Arrhenius law, induced by long-term memory, which we determine analytically. This correction, which we show can be quantitatively significant,  takes the form of a second effective energy barrier  $E'<E$ and captures the dependence of rare event kinetics on initial conditions, which is a hallmark of long-term memory. Altogether, our results    quantify  the impact of long-term memory on rare event kinetics, beyond Arrhenius law.
\end{abstract}

\maketitle
 
Many physical  and chemical processes are controlled by ``rare'' events, referring to events that are qualitatively unlikely, but nonetheless  important because their realization has exceptional consequences \cite{hanggi1990reaction,pollak2005reaction}. Such events are ubiquitous in the context of chemical physics, as exemplified at the molecular scale by the formation or rupture of   bonds \cite{hanggi1990reaction} (e.g. in force spectroscopy experiments \cite{bullerjahn2014theory,bullerjahn2020non,bullerjahn2016analytical} or adhesion kinetics \cite{jeppesen2001impact}), protein folding \cite{ayaz2021non}, molecular motor dynamics~\cite{Badoual2002a,guerin2011,Guerin2011c}, or more generally nucleation events. Rare events are also relevant in other contexts, such as stock market crashes~\cite{Bouchaud1998} or climate~\cite{ragone2018computation} or population~\cite{kamenev2008colored,dykman2008disease} dynamics.  The kinetics of such events, quantified by the first-passage time  (FPT) to a target configuration, generally follows Arrhenius (also called Kramers, or Eyring-Kramers) law: the mean waiting  time for a rare event  is exponentially large with the energy barrier that has to be crossed to reach the target configuration \cite{hanggi1990reaction}. This picture is also valid in non-equilibrium systems with the definition of a pseudo-potential \cite{Freidlin1984,Maier1992,bouchet2016generalisation,delacruz2018minimum}. 
In the weak-noise limit, the mean FPT is generally obtained by analyzing the dynamics at the top of the (pseudo-)potential barrier, by expanding around the most probable path  leading to the target configuration. In this limit  the waiting time for a   rare event becomes larger than all relaxation times of the dynamics, and is thus independent of initial conditions. 

While the effect of memory on  first passage~\cite{ReviewBray,metzler2014first,lindenberg2019chemical,Sokolov2003,Likthman2006,guerin2016mean,delorme2015maximum} and rare event kinetics~\cite{ferrer2021fluid,ginot2022barrier,lavacchi2020barrier,lavacchi2022non,bullerjahn2020non,kappler2018memory,Caraglio2018,carlon2018effect,medina2018transition,goychuk2007anomalous,Sliusarenko2010,arutkin2020extreme,levernier2020kinetics,delorme2017pickands,goswami2023effects} has been the object of recent studies, an important open question arises as to whether Arrhenius law is still valid for stochastic processes (or ``reaction coordinates'') $x(t)$ displaying \textit{infinite} relaxation times, i.e. with correlation functions  decaying as a power-law rather than exponentially:  
\begin{equation}
 \phi(\tau)\equiv \lim_{t\to\infty}\frac{\langle x(t)x(t+\tau)\rangle}{\langle x^2(t)\rangle }\underset{\tau\to\infty}{\simeq} \frac{A}{\tau^{\alpha}}\label{llm},
\end{equation}
where $A>0$, $\alpha>0$ and $\langle x(t)\rangle=0$ by convention. 
Stochastic processes possessing the property (\ref{llm}) will  be called hereafter  \textit{long-term memory} processes~\cite{santhanam2008return,bunde2005long} and arise when their dynamics results from the evolution of an infinite number of degrees of freedom. 
Examples of    processes with long-term memory are provided by  the dynamics of polymers~\cite{Panja2010}, proteins~\cite{kou2004generalized,Min2005} or
interfaces \cite{ReviewBray},  but also  earthquakes \cite{lennartz2008long} or rainfalls \cite{bunde2013there}. It is known that long-term memory induces dispersed kinetics \cite{min2006kramers,goychuk2009viscoelastic} and correlations between successive realizations of rare events~\cite{eichner2007statistics,santhanam2008return,bunde2005long}; its impact on the kinetics of rare events however remains to be elucidated.  
In fact, this question was considered in Ref.~\cite{goychuk2007anomalous} by means of a generalized Fokker-Planck equation,  a controversial \cite{singh2019comment,bullerjahn2017unified,bullerjahn2020non} method which leads to the  
notable prediction that the mean FPT to a rare configuration is  infinite for a class of processes with long-term memory ; in Ref.~\cite{Sliusarenko2010}, it was noted that the standard so-called ``Wilemski-Fixman'' approximation \cite{WILEMSKI1974a} also predicts infinite mean FPTs  [when the exponent $\alpha$ defined in (\ref{llm}) satisfies $\alpha<1$] \cite{NoteRef28}. 
Nevertheless, these predictions of infinite mean FPTs for processes with long-term memory seem inconsistent with   numerical simulations \cite{singh2019comment,Sliusarenko2010,bullerjahn2017unified} and the mathematical results of Refs.~\cite{newell1962zero,pickands1969upcrossing,pickands1969asymptotic}, which  point to  finite mean FPTs. 
Such contradiction shows that the above mentioned  methods cannot be used to analyse the impact of long-term memory on rare event kinetics. 

\begin{figure}[t!]
	\includegraphics[width=8cm]{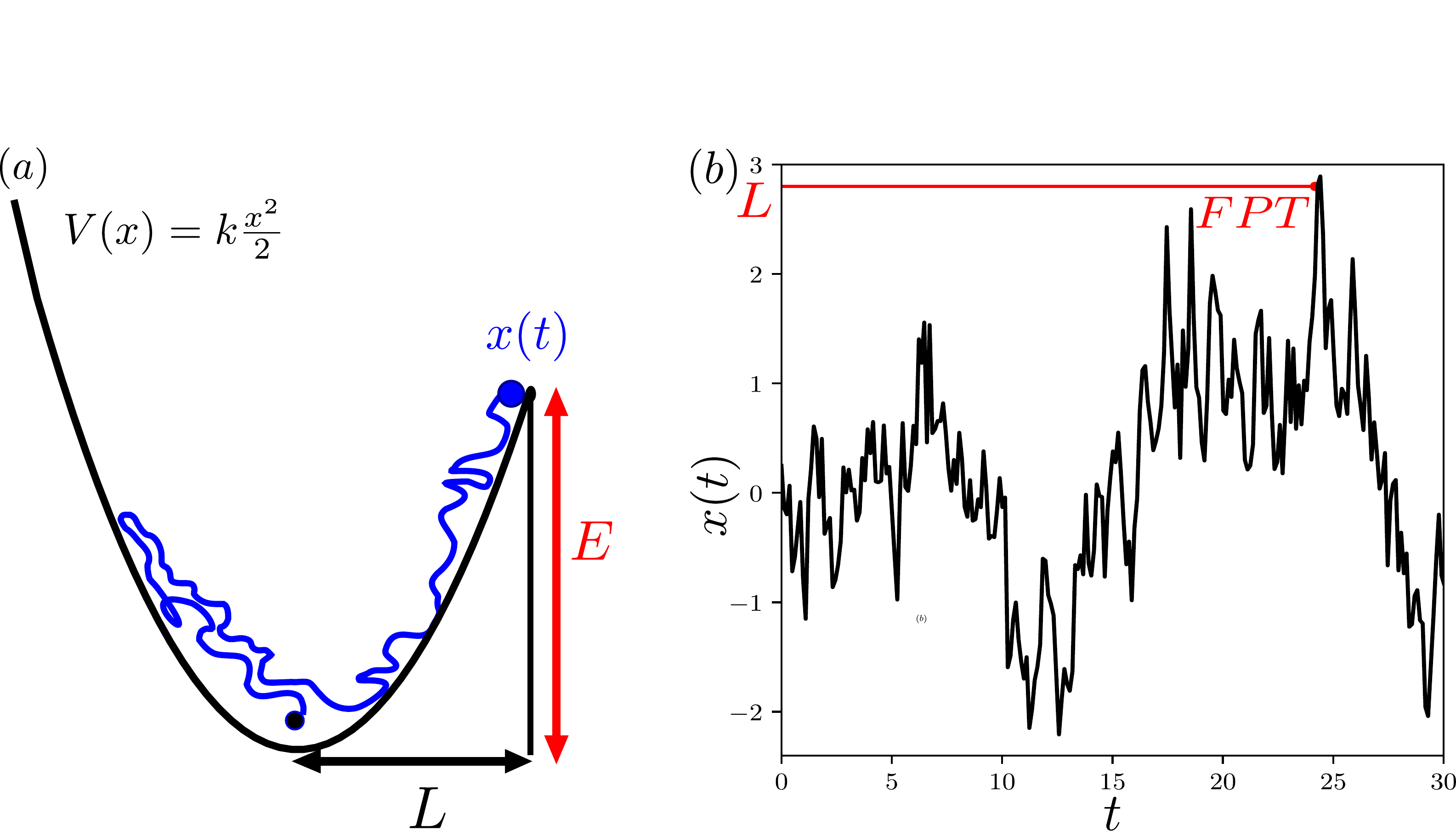}
	\caption{(a) Sketch of the problem. Let $x(t)$ be a random walker in a potential at temperature $\mathcal{T}$, submitted to a power-law friction kernel. In this example of long-term memory (meaning that the correlation function of $x(t)$ decay as a power-law), what is the mean FPT to a target at $x=L$ that can be reached only by overcoming an energy barrier  $E=V(L)-V(0)$ ?  (b) Sketch of the FPT for a single stochastic trajectory of $x(t)$.  }
	\label{FigSketch}
\end{figure}

Here, on the basis of a simple model of a  particle in a potential $V(x)$ at finite temperature with retarded friction force, we resolve this issue and quantify the impact of long-term memory on the kinetics of rare events. We generalize  to processes with long-term memory a  formalism  that was so far  restricted  to the analysis of either FPTs  in large confining volumes with flat energy landscapes \cite{guerin2016mean}, or of rare events without long-term memory \cite{levernier2020kinetics}. 
Our theory predicts finite mean FPTs, and is supported quantitatively by  numerical simulations. In the limit of large energy barriers -- called hereafter rare events limit, we show that Arrhenius law does hold, with however sub-exponential corrections induced by  the long-term memory,  which we determine explicitly. We find that long term memory  effectively induces  a second effective energy barrier of size $E'= E(1-\alpha)$ (for  $\alpha<1$), where $E=V(L)-V(0)$ is the size of the real barrier (see Fig. \ref{FigSketch}). 
We find that the prefactor of this correction, which we explicitly calculate, is much larger than the prefactor of the leading-order Arrhenius law, which implies that this correction is significant for a broad range of  energy barriers.

\vspace{0.4cm}
\textbf{Minimal model}\\ We consider a minimal  model of non-Markovian process $x(t)$ with long term memory at temperature $\mathcal{T}$, in a  confining potential that is assumed harmonic, see Fig.~\ref{FigSketch}(a). We  assume that  $x(t)$  obeys the overdamped Generalized Langevin Equation  (GLE) :
\begin{equation}
\int_0^t dt'\ K(  t-t'  ) \ \dot{x}(t')=- k \ x(t) + \xi(t) \label{GLE}.
\end{equation}
Here, the 1--dimensional random variable  $x(t)$ stands typically for the position of a particle, $K(t)$ represents the friction kernel, $k$ is the stiffness of the harmonic potential applied to the particle, and $\xi(t)$ is a Gaussian thermal force with zero mean whose magnitude is set by the fluctuation dissipation theorem $\langle \xi(t)\xi(t')\rangle=k_B\mathcal{T} K(\vert t-t'\vert)$. With these definitions the process $x(t)$ is Gaussian and  its stationary probability density function (pdf)  is   $p_s(x)=e^{-\frac{kx^2}{2k_B\mathcal{T}}}/\sqrt{2\pi l^2}$, where $l=\sqrt{k_B\mathcal{T}/k}$ is the confinement length.  Memory effects are encoded in the friction kernel $K(t)$, and  result typically from complex interactions of the variable $x(t)$ with other, potentially hidden,  degrees of freedom. The dynamics (\ref{GLE}) describes a variety of physical processes: (i) the motion of a tracer particle in a viscoelastic fluid \cite{mason1995optical,gisler1998tracer,mason1997particle}, 
(ii) the motion of a tagged particle attached to a polymer chain \cite{Panja2010,Panja2010a,bullerjahn2011monomer}, 
(iii) the dynamics of the distance between two protein residues as experimentally observed \cite{Min2005}. 
In the following we will mainly focus on scale invariant friction kernels: 
\begin{equation}
K(t)=\frac{K_\alpha}{\Gamma(1-\alpha)t^{\alpha}}\label{kernel},
\end{equation}
where  $0<\alpha<1$,  $K_\alpha$ is a transport  coefficient, and $\Gamma(\cdot)$ is the gamma function. While the theory presented below could be applied to other kernels, this choice \eqref{kernel}  is relevant to the physical examples  (i),(ii),(iii) above. 
Furthermore, in absence of target, the correlation function defined in (\ref{llm}) is $ \phi(t)=E_{\alpha}[-(t/\tau_d)^{\alpha}]$ \cite{goychuk2007anomalous,min2006kramers} where $\tau_d=(K_\alpha/k)^{1/\alpha}$  and $E_\alpha(\cdot)$ is the Mittag-Leffler function. Since $E_\alpha(-u){\sim}1/[\Gamma(1-\alpha)u] $ for large arguments, the choice of kernel (\ref{kernel}) ensures that the process $x(t)$ displays long-term memory as defined in \eqref{llm}: there is no finite relaxation time  in the correlation function, and   $A=K_\alpha/[\Gamma(1-\alpha) k]$ (SM, Section A).   

If one imposes the initial condition   $x(0)=x_0$, the average path $m_0(t)\equiv\langle x(t)\rangle_{x(0)=x_0}$ and the covariance $\sigma(t,t')\equiv\text{Cov}(x(t),x(t'))_{x(0)=x_0}$  conditional to $x(0)=x_0$  read~\cite{Eaton1983}
\begin{equation}
m_0(t)=x_0 \phi(t),\hspace{0.15cm}\sigma(t,t')=l^2[\phi(\vert t-t'\vert)-\phi(t)\phi(t')].\label{moments}
\end{equation} 
We also define $\psi(t)=\sigma(t,t)$ as  the Mean Squared Displacement (MSD) of  $x(t)$. In absence of potential ($k=0$), $x(t)$ is the fractional Brownian motion of Hurst exponent $H=\alpha/2$ ; for finite $k$ this regime is realized at short times, when the harmonic force is negligible, as seen from the  MSD:
\begin{equation}
 \psi(t)\underset{t\to0}{\simeq} \kappa\  t^{2H}, \ \kappa=\frac{2k_B\mathcal{T}}{K_\alpha\Gamma(1+\alpha)}, \ H=\frac{\alpha}{2}. 
\end{equation}
Hereafter we study the mean FPT of the process $x(t)$ defined by \eqref{GLE}, \eqref{kernel} to a target threshold $x=L$, with an initial configuration either drawn from the equilibrium distribution   or set by  $x(0)=x_0$. 

\begin{figure}[t!]
	\includegraphics[width=8.5cm]{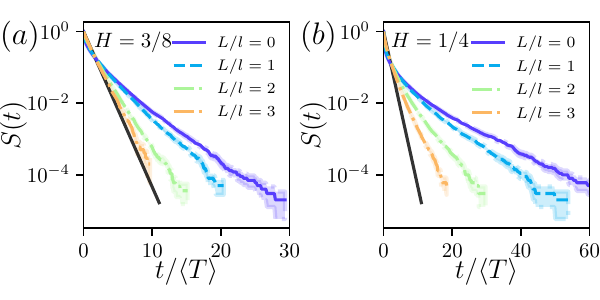}
	\caption{Survival probabilities $S(t)$ for (a) $H=3/8$ and (b) $H=1/4$, as measured in  numerical simulations. Here $x_0$ is drawn from the equilibrium distribution $p_s(x)$.  The black line represents $S(t)=e^{-t/\langle T\rangle}$. Error bars represent 68$\%$ confidence intervals, due to statistical uncertainties. 
	}
	\label{FigFPTDistr}
\end{figure}

\begin{figure}[t!]
	\includegraphics[width=8cm]{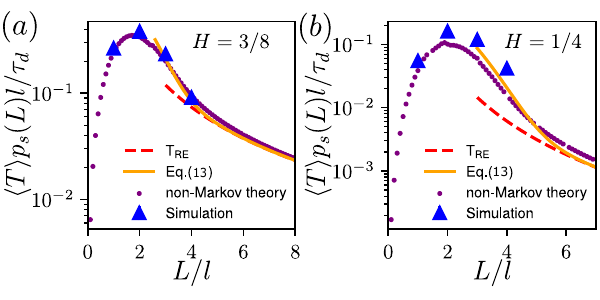}
	\caption{Mean FPT when the initial position is $x_0=0$ for (a) $H=3/8$ and (b) $H=1/4$. Symbols: numerical simulations; dots: numerical integration of Eqs.~(\ref{MFPT}, \ref{EqmPi}); dashed red line: Arrhenius law at leading order, Eq.~(\ref{TRE}); orange full line: refined Arrhenius law (\ref{TRefined}), including the corrections due to long-term memory. We have used the values $\nu_{3/8}=5.26$ and $\nu_{1/4}=5.0$ calculated in Ref.~\cite{levernier2020kinetics}.  
	}
	\label{FigMFPT}
\end{figure}

\vspace{0.4cm}
\textbf{Numerical analysis}\\ We have performed numerical simulations of the GLE (\ref{GLE}) by using a modified version of the circulant matrix algorithm \cite{davies1987tests} described in Ref.~\cite{dietrich_fast_1997}, which is an exact generator of  $x(t)$ at sampling times $t_n=n\times d t$  for any value of the time step $d t$. The used values of $dt$ are indicated in the Supplemental Material (SM, Section D) and are always smaller than $2\times10^{-5}\tau_d$. We used the two values of $H=\alpha/2$ that are used in classical polymer models : either a semi-flexible chain ($H=3/8$) or  a flexible (Rouse) chain  without hydrodynamic interactions ($H=1/4$). 
For each trajectory $\{x(t_n)\}$ we measured the FPT to $L$. The resulting survival probability $S(t)$ (defined as the probability that the FPT is larger than $t$) is shown in Fig.~\ref{FigFPTDistr}. Our numerical results are consistent with the mathematical results of Refs~\cite{newell1962zero,pickands1969upcrossing,pickands1969asymptotic}: an exponential decay of $S(t)$ in the rare events limit  $L\to\infty$, and  a stretched exponential behavior for $L=0$. This numerical analysis thus further supports that the mean FPT is finite (see Fig.~\ref{FigMFPT}).

\begin{figure}[t!]
	\includegraphics[width=8.5cm]{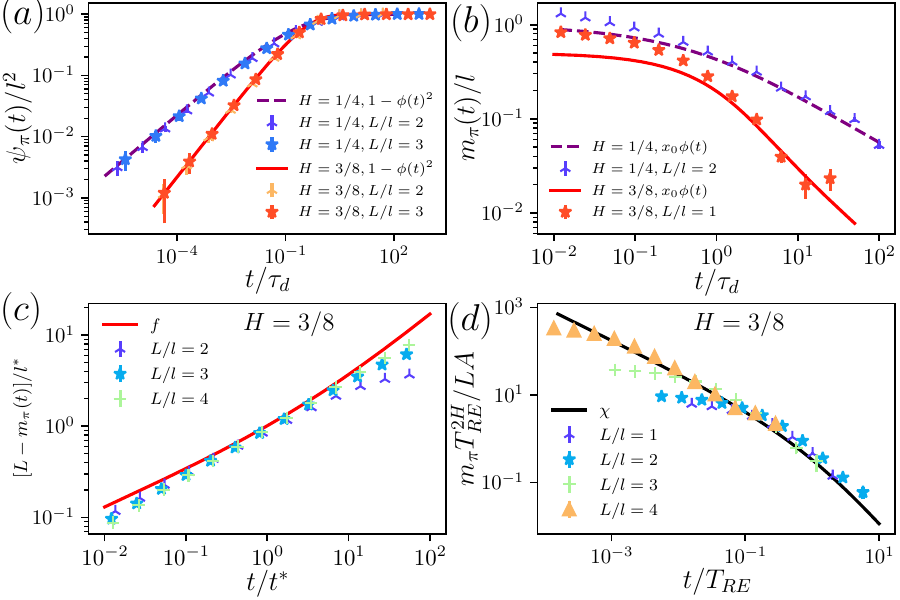}
	\caption{(a) Check of the stationary covariance approximation (i.e. $\sigma_\pi(t,t')\simeq\sigma(t,t')$): comparison between $\psi_\pi(t)=\text{Var}(x_\pi(t)) $ measured in numerical simulations (symbols) and $\psi(t)$ (dashed line: $H=3/8$, full line $H=1/4$). (b): Check of Eq.~(\ref{LongTimemPi}): comparison between the value $m_\pi(t)$ in simulations (symbols) and $x_0\phi(t)$ (full line:  $x_0=l/2$ for $H=3/8$; dashed line $x_0=l$ for $H=1/4$). 
Note that $m_\pi(t)\simeq x_0\phi(t)$ is expected at large times only.
	(c) Check of the short-time scaling regime   for $H=3/8$. 
	(d) Check of the long-time scaling regime  (\ref{mpi}) for $H=3/8$.  	In (a),(c),(d), the initial position is drawn from an equilibrium distribution, corresponding to our predictions for $x_0=0$. When present, error bars represent $68\%$ confidence intervals. }
	\label{FigControls}
\end{figure}

\vspace{0.4cm}
\textbf{General non-Markovian analysis } \\We now proceed to the theoretical determination of the mean FPT to $x=L$, denoted  $\langle T \rangle$, with fixed initial condition $x(0)=x_0$ [the case of stationary initial conditions can be obtained by averaging over $p_s(x_0)$].
 Our approach consists in generalizing the tools developed in Refs.~\cite{levernier2020kinetics,guerin2016mean,levernier2019survival}, which, in the context of rare event kinetics,  have been used so far only to analyze processes with a \textit{finite} maximal relaxation time \cite{levernier2020kinetics}. We describe the main steps of the approach for completeness ; details can be found in  SM (Section B). 
We start with the following general exact expression of the mean FPT, derived in Ref.~\cite{guerin2016mean}:
\begin{equation}
\langle T \rangle p_s(L) = \int_0^\infty dt \ [p_\pi(L,t) - p(L,t)], \label{MFPT}
\end{equation}
where we have introduced $p_\pi(x,t)$ as the  pdf  of the  process  $x_\pi(t)\equiv x(t+T)$, where $T$ is the FPT; $x_\pi(t)$ is thus the process after a first-passage event. 
To characterize  $p_\pi(x,t)$, we assume that the process $x_\pi(t)$ is Gaussian (as is $x(t)$), and thus fully characterized by its first moment  $m_\pi(t)=\langle x_\pi(t)\rangle$, and covariance $\sigma_\pi(t,t')\simeq \sigma(t,t')$ that is assumed to be identical to that of the unconditioned process $x(t)$. The validity of these hypotheses has been checked numerically [Fig.~\ref{FigControls}(a) and SM, Section D] and  analytically for weakly non-Markovian processes (SM, Sections E). With these approximations,  Eq.~\eqref{MFPT} becomes
\begin{equation}
\langle T \rangle p_s(L) = \int_0^\infty \frac{dt}{\sqrt{2\pi\psi(t)}} \left[ e^{-\frac{[m_\pi(t)-L]^2}{2\psi(t)}} - e^{-\frac{[x_0\phi(t)-L]^2}{2\psi(t)}}\right]\label{MFPT2}.
\end{equation}
The so-far unknown quantity $m_\pi(t)$ can then be determined self-consistently by analyzing a generalized version of the renewal equation (see SM, Section B), leading to 
\begin{align}
\int_0^\infty dt  \Bigg\{ & \frac{e^{-\frac{[m_\pi(t)-L]^2}{2\psi(t)}}}{\sqrt{2\pi\psi(t)}} \left(m_\pi(t+\tau) -\left[m_\pi(t)-L\right]\frac{\sigma(t+\tau,t)}{\sigma(t,t)}-L\phi(\tau)\right) \nonumber\\
-& \frac{e^{-\frac{[x_0\phi(t)-L]^2}{2\psi(t)}}}{\sqrt{2\pi\psi(t)}} \left(x_0\phi(t+\tau) -\left[x_0\phi(t)-L\right]\frac{\sigma(t+\tau,t)}{\sigma(t,t)}-L\phi(\tau)\right) \Bigg\}=0. \label{EqmPi}
\end{align} 
This equation generalizes  similar equations in Refs.~\cite{guerin2016mean,levernier2020kinetics}, which were restricred on the determination of $p_\pi(L,t)$ at short times and  thus did not enable the analysis of long-term memory effects. This integral equation, together with the condition $m_\pi(0)=L$,   allows to determine the only unknown $m_\pi(t)$ :  this finally  gives access to $\langle T \rangle$ thanks to Eq.~(\ref{MFPT2}). 

\vspace{0.4cm}
\textbf{General results}\\ This approach first shows unambiguously that the mean FPT is finite. Indeed, we show in SM (Section B) 
that the solution to Eq.~(\ref{EqmPi})  satisfies at long times
\begin{equation}
m_\pi(t)\underset{t\to\infty}{\simeq} x_0 \ \phi(t)\label{LongTimemPi},
\end{equation}
which can be checked directly in numerical simulations, see figure \ref{FigControls}(b). This scaling,  together with Eq.~(\ref{MFPT}), shows that the mean FPT is finite. 
This contradicts the results obtained with the generalized Fokker-Planck equation  \cite{goychuk2007anomalous} or with the  Wilemski-Fixman approximation \cite{WILEMSKI1974a}. The latter   amounts to assuming that the process is at all times  in an equilibrium state, and would thus yield    $m_\pi(t)\simeq L\phi(t)$, leading to an infinite mean FPT when $\alpha<1$ (as noted earlier in a similar, but out of equilibrium, situation \cite{Sliusarenko2010}). Beyond this  proof of finiteness, our approach  yields a quantitative determination of $\langle T\rangle$ by solving numerically the integral equation (\ref{EqmPi}) for $m_\pi(t)$ and next using   Eq.~(\ref{MFPT}) ;  this shows  quantitative agreement with numerical simulations in Fig.~\ref{FigMFPT}.  

\vspace{0.4cm}
\textbf{Rare events limit $L\to\infty$}\\ We now consider the rare event limit to determine explicitly the impact of long-term memory on rare events kinetics. 
The  mean FPT  obtained by the method of matched asymptotics  which we sketch here; calculation details are provided in SM (Section C). 
The dynamics involves different time and length  scales ; two can be readily identified: (i) the confinement length $l$ and (ii) the length $l^*=k_B\mathcal{T}/F$, where $F=kL$ is the slope of the potential at $L$. The associated time scales are respectively (i) $\tau_d$ and (ii) the time $t^*$ at which the characteristic fluctuations $\sqrt{\kappa} (t^*)^{H}$ of the trajectories near the target become comparable to $l^*$, this leads to $t^*=(l^*/\sqrt{\kappa})^{1/H}$. Note that in the rare events limit $t^*\ll \tau_d$. 

The leading order term $T_{\text{RE}}$ of $\langle T\rangle$ in the $L\to\infty$ limit results from the contribution of timescales  $t\sim t^*\ll \tau_d$ only in \eqref{MFPT}. Indeed, after a time $t\gg t^*$, a particle initially at $L$ has typically moved  away from the target, so that $p_\pi(L,t)$ is exponentially small, whereas $p_\pi$ is of order 1 at very short times $t\sim t^*$.  In turn, if the starting position is typically not close from $L$, $p(L,t)$ is exponentially small with $L$ at all times. 
The above consideration suggests to look for solutions of the form $ m_\pi(t)\simeq L- l^* f(t/t^*)$ ; inserting this ansatz in \eqref{EqmPi} and taking the rare event limit leads to an equation for $f$ that depends only on $H$, justifying our ansatz. The mean FPT at leading order is then obtained as
\begin{equation}
\langle T\rangle \underset{L\to\infty}{\sim}   \frac{l^{\frac{2}{H}-1}\ \nu_H}{L^{\frac{1}{H}-1}\ \kappa^{\frac{1}{2H} }}  \times e^{ \beta E} \equiv  T_{\text{RE}}, \label{TRE} \end{equation}
where $ \nu_H = \int_0^\infty \frac{du}{u^H} \ e^{-f^2(u)/2u^{2H}} $ depends only on $H$, $E=kL^2/2$ is the energy barrier and $\beta=1/(k_B\mathcal{T})$.  
 This leading order result displays the usual Arrhenius factor $e^{\beta E}$, which is the hallmark  of rare event kinetics, and  is compatible with the mathematical results of Pickands \cite{pickands1969upcrossing}.  Of note, it is controlled only by the short time behavior of the MSD $\psi(t)$, and is independent of the long time relaxation of correlations, and thus of long-term memory. It is indeed identical to the results of Ref.~\cite{levernier2020kinetics} obtained for non Markovian processes with the same MSD at short times but finite relaxation time. To prove this result self-consistently, we need to estimate the contributions to $\langle T\rangle$ in \eqref{MFPT}, that are induced by the behaviour of the integrand at time scales $t\gg t^*$.  These contributions are expected to be relevant in the case of  long-term memory, due to the slow decay of correlation functions.

Here, the key point is to note that, in addition to the previously identified timescales $\tau_d$ and $t^*$, a third relevant timescale for the dynamics of $x_\pi(t)$ is the time $T_{\mathrm{RE}}$ it-self. Indeed, we show in SM that 
$m_\pi(t,L)$   can be written    for $t\gg t^*$  :
\begin{equation}
m_\pi(t)\simeq 
\begin{cases}
L\ \phi_\pi(t) &  (t^*\ll t =\mathcal{O}(\tau_d)\ll T_{\text{RE}}) \\
\frac{LA}{T_{\text{RE}}^{\alpha}} \chi \left(\frac{t}{T_{\text{RE}}}\right) &  (\tau_d \ll t=\mathcal{O}(T_{\text{RE}}))\label{mpi}
\end{cases}, 
\end{equation}
where $A$ is defined in \eqref{llm} and $\chi$ and $\phi_\pi$ are  scaling functions. 
The analysis of  Eq.~\eqref{EqmPi} at   timescales $\tau_d$ and $T_{\text{RE}}$, respectively, enables us to obtain equations for $\phi_\pi$ and $\chi$ that can be solved, leading to
\begin{equation}
\phi_\pi (t)=\phi(t), \hspace{0.3cm}\chi(y)=\alpha  \left(1-\frac{x_0}{L}\right)\Gamma(-\alpha,y)e^{y}+ \frac{ x_0}{y^{\alpha}L}, 
\end{equation}
where $\Gamma(s,y)=\int_y^\infty t^{s-1}e^{-t}dt $ is the upper incomplete gamma function.  
Finally,  inserting the  scaling forms for $m_\pi(t)$ 
into Eq.~(\ref{MFPT}), we obtain 
\begin{equation}
\langle T\rangle \simeq T_{\text{RE}} + T_{\text{RE}} ^{1-\alpha} \times \frac{A}{l^2} L(L-x_0)\Gamma(1-\alpha) \label{TRefined}.
\end{equation}
This is the central result of this Letter. It  confirms the validity of the leading order term $T_{\text{RE}}$, which is independent of long-term memory, and explicitly determines the subleading term, which is induced by long-term memory, as seen by the factor $A$  that characterizes the long-time decay of fluctuations.  Several remarks are in order: 
(i) Since $T_{\text{RE}}\propto e^{\beta  E}$,  the correction due to long-term memory is of order $e^{\beta E'}$ with an effective energy barrier $  E'=   E (1-\alpha)$. The smaller the value of $\alpha$ the larger the value of $ E'$, so that the convergence to the rare event limit is expected to be slower for small $\alpha$ (where  non-Markovian effects are stronger).
 (ii) Furthermore, the pre-exponential factor is clearly much larger for the corrective term than for the leading order term in the limit $L\to\infty$, so that the corrective term can be quantitatively important. Indeed, as observed in  figure \ref{FigMFPT}, taking into account this correction is essential to predict the rare event kinetics for not-too-large values of $L$. 
(iii) Eq. (\ref{TRefined}) shows that the subleading correction depends on the initial position $x_0$: because of long-term memory, initial conditions can thus impact quantitatively rare event kinetics. 
(iv) As a further validation of our analysis,  the expected scaling behaviors of $m_\pi$ are given in Fig. \ref{FigControls} (c),(d)  and hold in the large $L$ limit, with discrepancies at small times in Fig.~\ref{FigControls}(c) due to  limitations in the choice of the time step (see SM, Section D for additional parameters). 

\vspace{0.4cm}
\textbf{Conclusion} \\
We have proposed a theoretical analysis of the classical Kramers escape problem for non Markovian processes with long-term memory.  Although our approach is approximate, it captures the essence of memory effects and allows for a quantitative determination of the  mean FPT to a target, which we unambiguously show is finite,  whereas all existing theoretical approaches so far incorrectly predicted infinite mean FPTs  (for $\alpha<1$). This comes from the  assumption  of a system at equilibrium at all times that is  implicitly made in the methods that have been employed so far, namely the Wilemski-Fixman approximation or the generalized Fokker-Planck equation approach. Such hypothesis  is too strong to take properly into account long-term memory effects. In our approach, the genuine non-equilibriumness  of the system  upon a   first passage event  manifests itself in the  trajectory $m_\pi(t)$, whose behaviour at very long times is affected by long-term memory. 
In the rare event limit, we have explicitly determined the correction to Arrhenius laws, which is  due to long-term memory. This takes the form of a second effective energy barrier of size $E'=E(1-\alpha)$, which we show can be quantitatively significant, and captures the dependence of the kinetics on initial conditions.  
 It is known that Arrhenius laws can be identified for non-Gaussian models by considering the linearized dynamics around the target~\cite{levernier2020kinetics}. Since our study reveals that the effect of long-term memory on rare event kinetics comes from the slow dynamics at the bottom of the potential only, we may  expect that our main result (\ref{TRefined}) could be generalized to non-Gaussian models. Moreover, although we have focused here on a simple model of a particle with viscoelastic friction at equilibrium at constant temperature, it is clear that our arguments to identify the mean FPT could be adapted to active models where the fluctuation-dissipation theorem does not hold. Indeed, Eq.~(\ref{MFPT2}) and (\ref{EqmPi}) would still be valid, and would involve similarly  the  properties of the process in absence of target ($A, \phi,p_s,\kappa,...$), which are in principle still accessible from the definition of the process in Eq.~(\ref{GLE}), even if the fluctuation-dissipation relation does not hold because of       active effects  \cite{Sliusarenko2010} .
Last, because our approach puts forward  deviations from Arrhenius law due to long-term memory, we also anticipate deviations from exponential laws for the distribution of FPTs,  that could be studied by generalizing our approach to higher moments of the FPT, possibly giving access to the analytical study of extreme events clustering and dispersed kinetics.  Altogether, our results   shed light on the effect of long-term memory on rare event kinetics, beyond Arrhenius laws. 

 \begin{acknowledgments}
 T. G. acknowledges the support of the grant \textit{ComplexEncounters}, ANR-21-CE30-0020. Computer time for this study was provided by the computing facilities MCIA (Mesocentre de Calcul Intensif Aquitain) of the Universit\'e de Bordeaux and of the  Universit\'e de Pau et des Pays de l’Adour.
\end{acknowledgments} 


 \vspace{3cm}
 
 \begin{center}\Large{\textbf{Supplemental Material}}\end{center} 
 
\vspace{2cm}

 In this Supplemental Material, we provide
\begin{enumerate}
\item calculation details to obtain the solution of the GLE equation (without target) [Section \ref{SolutionGLE}]. 
\item a detailed derivation of the equations of the non-Markovian theory [Section \ref{DerivationEquations}]
\item calculation details for the asymptotic analysis in the rare event limit $L\to\infty$ [Section \ref{BL}]. 
\item Details on simulations and additional simulation data to check the Gaussian behavior of trajectories in the future of first passage events and our scaling arguments [Section \ref{FurtherChecks}]. 
\item  A note on the exactness of the approach for weakly non-Markovian processes [Section \ref{Perturbation}].
\end{enumerate}

\subsection{Solution of the Generalized Langevin Equation (without absorbing target)}
\label{SolutionGLE}
Here, we consider the dynamics given by the overdamped GLE
\begin{align}
&\int_0^t dt' K(\vert t-t'\vert ) \dot{x}(t')=- k \ x(t) + \xi(t), & \langle \xi(t)\xi(t')\rangle=k_B\mathcal{T} \ K(\vert t-t'\vert) \label{GLESi}. 
\end{align}
In absence of target, the solution of this equation is well known \cite{goychuk2007anomalous}, it is reminded here for the sake of completeness. Since the above equation is linear, the resulting process $x(t)$ is  Gaussian and is fully characterized by its two first moments. Denoting $\widetilde{f}(s) = \int_0^\infty f(t) e^{-st}dt$  the Laplace transform of a function $f$, we obtain
\begin{equation}\label{SiFree1}
\xtil(s) = \frac{ \xitil(s) + x(0) \tilde{K}(s)}{s\tilde{K}(s)+k}. 
\end{equation}
We also write
\begin{align}\label{SiFree2}
\langle \xitil(s) \xitil(s') \rangle &=k_B\mathcal{T}  \int_0^{\infty} dt \int_0^{\infty} dt' e^{-(st+s't')} K(|t-t'|) ,   \\ 
                 &= k_B\mathcal{T} \int_0^{\infty} dt' \int_0^{\infty}d\tau\ e^{-(s+s')t'-s\tau} K(\tau)  + k_B\mathcal{T} \int_0^{\infty} dt \int_0^{\infty}d\tau' \ e^{-(s+s')t-s'\tau'} K(\tau'), \label{Interm} \\
                 &=k_B\mathcal{T}  \frac{\tilde{K}(s) + \tilde{K}(s')}{s+s'},
\end{align}
where Eq.~(\ref{Interm}) has been obtained by setting $t=t'+\tau$ for $t>t'$ and $t'=t+\tau'$ for $t'>t$.  Using this result and (\ref{SiFree1}) yields, for an initially equilibrated initial position $\langle x(0)^2\rangle=k_B\mathcal{T}/k$:
\begin{align}\label{FJES}
\langle \xtil(s)\xtil(s')\rangle  &= \frac{ k_B\mathcal{T} }{[s\tilde{K}(s)+k][s'\tilde{K}(s')+k]}\left\{  \frac{\tilde{K}(s) + \tilde{K}(s')}{s+s'} + \frac{\tilde{K}(s)\tilde{K}(s')}{k} \right\},\nonumber\\
&= \frac{ k_B\mathcal{T} }{k (s+s')  }\left\{   \frac{\tilde{K}(s)}{s\tilde{K}(s)+k} +  \frac{\tilde{K}(s')}{s'\tilde{K}(s')+k}   \right\}.  
\end{align}
We may recognize that if one sets
\begin{align}
\langle x(t)x(t')\rangle = l^2 \phi(\vert t -t' \vert ),
\end{align}
then, using the same procedure as in Eq.~(\ref{Interm}), 
\begin{align}
\langle \tilde{x}(s)\tilde{x}(s')\rangle = \frac{l^2}{s+s'}[\tilde{\phi}(s)+ \tilde{\phi}(s')]. 
\end{align}
Comparing the above equation with  (\ref{FJES}) leads to
\begin{align}
&\tilde{\phi}(s)=\frac{\tilde{K}(s)}{s\tilde{K}(s)+k}. 
\end{align}
This formula is valid for arbitrary kernel. For the power-law kernel (3) of the main text, we  obtain $\tilde{K}=K_\alpha/s^{1-\alpha}$ and $\phi(t)$ is a Mittag-Leffler function:
\begin{align}
&\tilde{\phi}(s)=\frac{K_\alpha s^\alpha}{s[s^\alpha K_\alpha+k]},&\phi(t) = E_\alpha \left(-\left[\frac{t}{\tau_d}\right]^{\alpha} \right). 
\end{align}

The mean and covariance of the process when $x(0)=x_0$ is fixed can be obtained by using general formulas on conditional means and covariances for Gaussian processes, see e.g.  chapter 3 in Ref.~\cite{Eaton1983}:
\begin{align}
&\mathbb{E}(A\vert Y=y)=\mathbb{E}(A)-\frac{\text{Cov}(A,Y)}{\text{Var}(A)}(\mathbb{E}(Y)-y), \label{CondGauss1}\\ 
&\text{Cov}(A,B\vert Y=y)=\text{Cov}(A,B)-\frac{\text{Cov}(A,Y)\text{Cov}(B,Y)}{\text{Var}(A)}.\label{CondGauss2}
\end{align}
These formulas relate conditional averages and covariances to non-conditional ones, here $\mathbb{E}(A\vert Y=y)$ is the average of the variable $A$ given that the variable $Y$ takes the value $y$, and $\text{Cov}(A,B\vert Y=y)$ is the covariance of $A,B$ given that $Y=y$. Using these formulas, the average and the covariance of the process $x(t)$ conditional to $x(0)=x_0$ read
\begin{align}
&m_0(t)\equiv\mathbb{E}(x(t)\vert x(0)=x_0)=x_0 \phi(t), \label{IniMean}\\
&\sigma(t,t')\equiv\text{Cov}(x(t),x(t')\vert x(0)=0)=l^2[\phi(\vert t-t'\vert)-\phi(t)\phi(t')]. \label{IniCov}
\end{align}
We can also check these expressions by using directly (\ref{SiFree1}).

\subsection{Derivation of the equations  of the non-Markovian theory [Eqs.~(7,8,9,10,11)]}
\label{DerivationEquations}
Here we derive the equations that will give access to the mean first passage time (mean FPT) to $x=L$, when the stochastic process starts at $x_0$ at $t=0$. Let us start with a two-point generalized version of the renewal equation:
\begin{align}
&p(L,t;x_1,t+t_1)=\int_0^t dt' F(t') p(L,t;x_1,t+t_1\vert\text{FPT}=t')  \label{Ren2Points}. 
\end{align}
This exact equation comes from the fact that, if $x$ is observed at position $L$ at $t$, since the process is non-smooth, it means that $L$ was reached for the first time at some time $t'$, and the above equation is obtained by partitioning the event of observing $(L,x_1)$ at times $t,t+t_1$ over the value of the FPT.
Here, $p(L,t;x_1,t+t_1)$ is the joint probability density function (pdf)  of observing $x=L$ at time $t$ and the position $x=x_1$ at a later time $t+t_1$. The fact that the initial position is fixed is implicitly understood in this notation. Next, $p(L,t;x_1,t+t_1 \vert\text{FPT}=t')$ represents the probability density of observing $x=L$ at time $t$ and  $x=x_1$ at a later time $t+t_1$ given that the FPT is $t'$.  Note that, as originally noted in Ref.~\cite{Likthman2006}, for non-Markovian processes, it is necessary to keep the information that the target was reached at $t'$ for the first time in the propagators, this condition is different from the condition that $x(t')=L$ which would hold for Markovian processes.

 Now, we introduce the process in the future of the FPT, $x_\pi(t)\equiv x(t+\text{FPT})$ and we denote as $p_\pi(y,t)$ its pdf at time $t$ (after the FPT). By definition,
\begin{align}
p_\pi(L,t;x_1,t+t_1)=\int_0^\infty d\tau F(\tau)p(L,t+\tau;x_1,t+t_1+\tau \vert \text{FPT}=\tau) \label{Def2PointsPPi}.
\end{align}
We also define the stationary probability density of observing $x=L$ at some time and $x_1$ after a time $t_1$ has elapsed:
\begin{align}
p_s(L;x_1,t_1)\equiv \lim_{t\to\infty}p(L,t;x_1,t+t_1). 
\end{align}
We now consider Eq.~(\ref{Ren2Points}), where we substract $p_s(L;x_1,t_1)$ on both sides, leading to
\begin{align}
p(L,t;x_1,t+t_1)-p_s(L;x_1,t_1) =&\int_0^t dt' F(t') [p(L,t;x_1,t+t_1\text{FPT}=t')-p_s(L;x_1,t_1)] \nonumber\\
&- \int_t^\infty d\tau F(t')p_s(L;x_1,t_1)  \label{04381},
\end{align}
where we have used the fact that $\int_0^\infty dt F(t)=1$. To proceed further, we remark that 
\begin{align}
\int_0^\infty dt \int_t^\infty dt' F(t')=\int_0^\infty dt' \int_0^{t'} dt F(t')=\int_0^\infty dt' t' F(t')= \langle T\rangle \label{Trick1}.
\end{align}
We also note the following equalities:
\begin{align}
\int_0^\infty dt \int_0^t &dt' F(t')  [p(L,t;x_1,t+t_1\vert \text{FPT}=t')-p_s(L;x_1,t_1)] ,\nonumber\\
&=\int_0^\infty dt' \int_{t'}^\infty dt \ F(t') \ [p(L,t;x_1,t+t_1\vert \text{FPT}=t')-p_s(L;x_1,t_1)], \label{line1}\\
&=\int_0^\infty dt' \int_{0}^\infty du \ F(t') \ [p(L,t'+u;x_1,t'+t_1+u\vert  \text{FPT}=t')-p_s(L;x_1,t_1)], \label{line2}\\
&=\int_0^\infty du  \int_{0}^\infty dt' \ F(t') \ [p(L,t'+u;x_1,t'+t_1+u \vert \text{FPT}=t')-p_s(L;x_1,t_1)],\label{line3}\\
&=\int_0^\infty du \ [p_\pi(L,u;x_1,u+t_1)-p_s(L;x_1,t_1)]\label{Trick2},
\end{align}
where the successive calculation steps are: (i) the inversion of the order of integration  for the variables $(t,t')$ in Eq.~(\ref{line1}), (ii) the change of variable $t=u+t'$ in  Eq.~(\ref{line2}), (iii) again a change in the order of integration between the variables $u,t'$ in (\ref{line3}), and (iv) finally the use of the definition (\ref{Def2PointsPPi}) to simplify the integral. 
Next, using Eqs.~(\ref{Trick1}) and (\ref{Trick2}), we see that that integrating Eq.~(\ref{04381}) over $t$ leads to
\begin{align}
\int_0^\infty dt \ [p_\pi(L,t;x_1,t+t_1)-p(L,t;x_1,t+t_1)] =  \langle T\rangle p_s(L;x_1,t+t_1).  \label{GenEq}
\end{align}
This equation is general and exact, as soon as $p_s$ exists, for any continuous non-smooth stochastic process (even non-Gaussian).  
 Integrating over $x_1$ leads to a general expression for the mean FPT:
\begin{align}
\langle T\rangle p_s(L) = \int_0^\infty dt \ [p_\pi(L,t)-p(L,t)]. \label{GenMFPT}
\end{align}
Next, we write $p_\pi(L,t;x_1,t+t_1)=p_\pi(L,t)p_\pi(x_1,t+t_1\vert L,t)$ (this is Bayes' formula). Using this, multiplying Eq.~(\ref{GenEq}) by $x_1$ and integrating over $x_1$ yields
\begin{align}
\int_0^\infty dt [p_\pi(L,t) m_\pi^*(t+t_1\vert L,t)-p(L,t) m_0^*(t+t_1\vert L,t)] =  \langle T\rangle p_s(L)m_s^*( t_1\vert L,0),\label{05942}
\end{align}
where $m_\pi^*(t+t_1\vert L,t)$ is the conditional average of $x_\pi(t+t_1)$ given that $x_\pi(t)=L$, and (similarly) $m_0^*(t+t_1\vert L,t)$ is the conditional average of $x(t+t_1)$ given that $x(t)=L$. Finally, $m_s^*(t_1)$ is the average of $x(t_1)$ given that the system is equilibrated at $t=0$, with the condition $x(0)=L$.
Combining Eqs.~(\ref{GenMFPT}) and (\ref{05942}), we obtain
\begin{align}
\int_0^\infty dt \{p_\pi(L,t) [m_\pi^*(t+t_1\vert L,t)-m_s^*(t_1)]-p(L,t) [m_0(t+t_1\vert L,t)-m_s^*(t_1)]\} = 0.    \label{ClosureGen} 
\end{align}
To proceed further, we  assume that, in the future of the FPT, the process $x_\pi(t)$ is Gaussian, with a mean $m_\pi(t)$ and a covariance $\sigma_\pi(t,t')\simeq \sigma(t,t')$ that is approximated by the stationary covariance conditioned to $x=0$ at $t=0$. The next step consists in using the above equations as closure relations to determine the mean FPT.  

We now write explicit expressions for $m_\pi^*,m_0^*,m_s^*$. Using the general formula (\ref{CondGauss1}) for conditional averages, where we use $A=x_\pi(t)$, $Y=x_\pi(t+t_1)$ and $y=L$, we obtain
\begin{align}
&m_\pi^*(t+t_1\vert L,t)=m_\pi(t+t_1)-\frac{\sigma(t+t_1,t)}{\psi(t)}[m_\pi(t)-L], 
\end{align}
where $\psi(t)=\sigma(t,t)=l^2[1-\phi(t)^2]$ is the mean square displacement of the process $x(t)$ conditioned to $x(0)=0$. Similarly, applying again Eq.~(\ref{CondGauss1}) for $A=x(t)$, $Y=x(t+t_1)$ and $y=L$, we obtain
\begin{align}
&m_0^*(t+t_1\vert L,t)=m_0(t+t_1)-\frac{\sigma(t+t_1,t)}{\psi(t)}[m_0(t)-L]. 
\end{align}
Taking the limit $t\to\infty$ in the above formula enables us to identify $m_s^*$: 
\begin{align}
&m_s^*(t_1\vert L,0)=L \phi(t_1).
\end{align}
We also note that, for Gaussian propagators, 
\begin{align}
&p_\pi(L,t)=\frac{e^{-[L-m_\pi(t)]^2/2\psi(t)}}{\sqrt{2\pi\psi(t)}},& p(L,t)=\frac{e^{-[L-m_0(t)]^2/2\psi(t)}}{\sqrt{2\pi\psi(t)}}. 
\end{align}
Collecting these results, the closure equation (\ref{ClosureGen}) for $m_\pi(t)$ becomes 
\begin{align}
\mathcal{H}(\tau)\equiv &\int_0^\infty dt\Bigg\{ \frac{e^{-[L-m_\pi(t)]^2/2\psi(t)}}{[\psi(t)]^{1/2}}\left[m_\pi(t+\tau)-[m_\pi(t)-L]\frac{\phi(\tau)-\phi(t)\phi(t+\tau)}{1-\phi^2(t)}-L\phi(\tau)\right]\nonumber\\
&-\frac{e^{-[L-x_0\phi(t)]^2/2\psi(t)}}{[\psi(t)]^{1/2}}\left[x_0\phi(t+\tau)-[x_0\phi(t)-L]\frac{\phi(\tau)-\phi(t)\phi(t+\tau)}{1-\phi^2(t)}-L\phi(\tau)\right]\Bigg\}=0,\label{94323}
\end{align} 
and the expression (\ref{GenMFPT}) for the mean FPT becomes
\begin{align}
\langle T\rangle p_s(L)=\int_0^\infty dt\left\{ \frac{e^{-[L-m_\pi(t)]^2/2\psi(t)}}{[2\pi\psi(t)]^{1/2}}-\frac{e^{-(L-x_0\phi(t))^2/2\psi(t)}}{[2\pi\psi(t)]^{1/2}}\right\}\label{MFPTx0}.
\end{align} 
  
\subsubsection*{Behavior of $m_\pi(t)$ at large times and consequence for the mean FPT}
We note that, for large times, $\phi(t)$ becomes a small quantity for large times. Then we see that the second line of the integrande in Eq. (\ref{94323}) behaves as 
\begin{align}
\frac{e^{-(L-x_0\phi(t))^2/2\psi(t)}}{[2\pi\psi(t)]^{1/2}}\left[x_0\phi(t+\tau)-[x_0\phi(t)-L]\frac{\phi(\tau)-\phi(t)\phi(t+\tau)}{1-\phi^2(t)}-L\phi(\tau)\right] \underset{t\to\infty}{\simeq} x_0 (1-\phi(\tau))\phi(t).
\end{align}
Since $\phi(t)\simeq A/t^{2H}$ and $H<1/2$, we see that these terms have to be compensated so that the integral (\ref{94323}) exists; this implies that
\begin{align}
m_\pi(t) \underset{t\to\infty}{\simeq} x_0 \ \phi(t)\label{LongTimemPi},
\end{align}
and this equality should hold at all orders of $t^{-a}$ with $a<1$. 
If the behavior (\ref{LongTimemPi}) holds then  the mean FPT predicted by Eq.~(\ref{MFPTx0}) is finite.

\subsection{Asymptotic analysis in the rare event limit, $L\to\infty$} 
\label{BL}
Here, we  analyze the structure of the solution $m_\pi(t,L)$ in the limit $L\to\infty$. As mentioned in the main text, a natural length scale for the dynamics near the top of the potential is $l^*=k_B\mathcal{T}/F$, where $F=kL$ is the slope of the potential. Hence $l^*=l^2/L$. The associated time scale $t^*$ is the time at which $\psi(t^*)$ is of order $l^*$, this leads to $t^*=(l^*/\sqrt{\kappa})^{1/H}$. This suggests the ansatz 
\begin{align}
m_\pi(t,L)\simeq L-\frac{l^2}{L}f(t/t^*) , \hspace{1cm} t^*=\left(\frac{l^2}{L\sqrt{\kappa}}\right)^{1/H}. \label{BehaviormPiShortTimes}
\end{align}
Note that $t^*\to0$ when $L\to\infty$. Here $f$ is a scaling function that is determined by requiring that  $\mathcal{H}(\tau=t^*v)$, where $\mathcal{H}$ is defined in Eq.~(\ref{94323}), vanishes in the limit $L\to\infty$ (at fixed $v$):
\begin{align}
\mathcal{H}( t^*v) \underset{L\to\infty}{\simeq} \frac{l^2(t^*)^{1-H}}{L\sqrt{\kappa} } \int_0^\infty \frac{du}{u^H} e^{-\frac{f^2(u)}{2u^{2H}}} \left[-f(u+v)+f(u)\frac{u^{2H}+(u+v)^{2H}-v^{2H}}{2u^{2H}}+\frac{v^{2H}}{2}\right]=0, \label{Eq_for_f}
\end{align} 
where we have used $\phi(\tau)\simeq 1-\tau^{2H}\kappa/(2l^2)$ for small $\tau$ (so that $\psi(\tau)\simeq \kappa \tau^{2H}$). 
Solving this equation yields the scaling function $f$. Next, we investigate the behavior of $m_\pi(t)$ at time scales larger than $t^*$. 
It is   natural to assume that $m_\pi(t)$ admits a regime that varies at the same time scale $\tau_d$ as the original dynamics for $x(t)$, which leads us to the ansatz
\begin{align}
m_\pi(t,L)\simeq 
\begin{cases}
L-\frac{l^2}{L}f(t/t^*) & t=\mathcal{O}(t^*), \ (t\ll \tau_d), \\
L\ \phi_\pi(t) &t=\mathcal{O}(\tau_d), \ (t\gg t^*),
\end{cases} \label{TwoTImes}
\end{align}
where $\phi_\pi$ is a scaling function that is independent of $L$. The linear term in $L$ in factor of $\phi_\pi$ is justified by the fact that the matching with the solution at scale $t^*$ can be achieved with the conditions
\begin{align}
&\phi_\pi(t)\underset{t\to0}{\simeq} 1- c\ \frac{\kappa \ t^{2H}}{l^2}  , & f(u\to\infty)\simeq  c\ u^{2H},
\end{align}
where $c$ is a numerical constant. The equation for $\phi_\pi$ is obtained by looking at the behavior of $\mathcal{H}(\tau)$ for $L\to\infty$ at fixed $\tau$, the integrals can in fact be evaluated at times $t^*$ (all other terms are exponentially small) so that we obtain
\begin{align}
\mathcal{H}(\tau)\underset{L\to\infty}{\simeq} \frac{(t^*)^{1-H} }{\sqrt{\kappa}} \int_0^\infty du   \frac{e^{-f^2(u)/2u^{2H}}}{u^H}\left[L\ \phi_\pi(\tau)-L\ \phi(\tau)\right] . 
\end{align} 
Since $\mathcal{H}(\tau)$ has to vanish for all $\tau$ we conclude that $\phi_\pi=\phi$: thus at this time scale $\tau_d$ the average trajectory in the future of the FPT is, at leading order, the same as the trajectory constrained to $x(0)=L$ starting from an equilibrium configuration. However, it is obvious that this behavior (\ref{TwoTImes}) cannot hold at very long times since at this stage it is not possible to connect the long-time behavior of $m_\pi = L\phi(t)$ to the already identified behavior given by Eq.~(\ref{LongTimemPi}), where $m_\pi(t)\simeq x_0 \phi(t)$. Hence, we have to postulate the existence of at least one additional  longer time scales. 
Let us define now $T_{\text{RE}}$ as  
\begin{align}
&T_{\text{RE}} = e^{  L^2/(2l^2)} \frac{(t^*)^{1-H} l  }{ \kappa^{1/2}}\nu_H, &\ \nu_H= \int_0^{\infty} du\ \frac{e^{-f^2(u)/2u^{2H}}}{u^H}. \label{DefTRE}
\end{align}
It turns out that $T_{\text{RE}}$ will be the value of the mean FPT at leading order when $L\to\infty$, but since this is not obvious for our long-term memory process we use the above equation as a definition for $T_{\text{RE}}$. Note that $L^2/(2l^2)=  E/k_B\mathcal{T}$ is the value of the energy barrier to be crossed to reach the target point. 
Anticipating   the final result, we postulate that $T_{\text{RE}}$ is also a characteristic time scale for $m_\pi$. Considering this third time scale, the behavior of $m_\pi$ reads
\begin{align}
m_\pi(t,L)\simeq 
\begin{cases}
L-\frac{l^2}{L}f(t/t^*) & t=\mathcal{O}(t^*), \ (t\ll \tau_d), \\
L\ \phi(t) &t=\mathcal{O}(\tau_d), \ (t^*\ll t \ll T_{\text{RE}}), \\
\frac{LA}{T_{\text{RE}}^{2H}} \chi \left(\frac{t}{T_{\text{RE}}}\right) &t=\mathcal{O}(T_{\text{RE}}), \ (\tau_d \ll t)
\end{cases} \label{ThreeTimes}
\end{align}
where $\chi$ is a scaling function. The term $LA/T_{\text{RE}}^{2H}$ in factor of $\chi$ is justified by the fact that the solutions at scales $\tau_d$ and $T_{\text{RE}}$ are matched (i.e. predict the same value for $m_\pi$) at the condition
\begin{align}
\chi(u)\underset{u\to0}{\simeq }1/u^{2H}\label{BehaviorChi1SmallArg}. 
\end{align}
We  find the equation for $\chi$ by calculating $\mathcal{H}(\tau=\overline{\tau}T_{\text{RE}})$ when $L\to\infty$ at fixed $\overline{\tau}$. The key remark is that since  $T_{\text{RE}}$ is  exponentially large with $L$,   the integral  (\ref{94323}) has  two contributions: a first one coming from $\tau$ of oder $t^*$ and a second one coming from $\tau=O(T_{\text{RE}})$. 
We note that
\begin{align}
\frac{\phi(\tau)-\phi(t+\tau)\phi(t)}{1-\phi^2(t)}\underset{t\ll \tau_d \ll\tau}{\simeq} \frac{A}{2\tau^{2H}},
\end{align}
so that, with $\tau=T_{\text{RE}}\overline{\tau}$ and $t=ut^*$, we have
\begin{align}
&m_\pi(t+\tau)=m_\pi(ut^*+\overline{\tau}T_{\text{RE}} ) \simeq \frac{LA}{T_{\text{RE}}^{2H} }\chi(\overline{\tau}),\\
&(m_\pi(t)-L)\frac{\phi(\tau)-\phi(t+\tau)\phi(t)}{1-\phi^2(t)}=  -\frac{l^2\ f(u)}{L\ T_{\text{RE}}^{2H}}\frac{A}{2\overline{\tau}^{2H}}\ll m_\pi(t+\tau). 
\end{align}
Following these considerations, we evaluate 
\begin{align}
\mathcal{H}(\tau=T_{\text{RE}}\overline{\tau})\underset{L\to\infty}{\simeq} \frac{(t^*)^{1-H}}{\sqrt{\kappa}} \frac{LA}{T_{\text{RE}}^{2H}} \int_0^\infty du \frac{e^{-f^2(u)/2u^{2H}}}{u^H}\left[\chi(\overline{\tau})-\frac{1}{ \overline{\tau}^{2H}}\right]\nonumber\\
+ \frac{LA\ T_{\text{RE}} }{l\  T_{\text{RE}}^{2H}} \int_0^\infty d\overline{t}\ e^{-\frac{L^2}{2l^2}} \left[ \chi (\overline{t}+\overline{\tau})-\frac{\tilde{x}_0}{(\overline{t}+\overline{\tau})^{2H}}\right],
\end{align} 
where $\tilde{x}_0=x_0/L$ and one keeps $\tilde{x}_0$ constant when taking the limit $L\to\infty$.  
Equating this expression to zero and using the definition of $T_{\text{RE}}$ in Eq.~(\ref{DefTRE}) we thus obtain
\begin{align}
    \chi(\overline{\tau})-\frac{1}{\overline{\tau}^{2H}} 
+ \int_0^\infty d\overline{t}\ \left[ \chi (\overline{t}+\overline{\tau})-\frac{\tilde{x}_0}{(\overline{t}+\overline{\tau})^{2H}}\right] =0 \label{EqChi1} .
\end{align} 
This equation can be solved by setting $G(\overline{\tau})=\chi(\overline{\tau})-\tilde{x}_0/\overline{\tau}^{2H}$, and differentiating with respect to $\overline{\tau} $:
\begin{align}
 G'(\overline{\tau})+(1-\tilde{x}_0)\frac{2H}{\overline{\tau}^{2H+1}} -G(\overline{\tau})
 = 0 ,
\end{align} 
where one has assumed that $G(\infty)=0$. The only solution that does not diverge exponentially for large arguments is 
\begin{align}
&G(\overline{\tau})=(1-\tilde{x}_0) 2H \ \Gamma(-2H,\overline{\tau})e^{\overline{\tau}},& \ \chi(\overline{\tau})=(1-\tilde{x}_0) 2H \ \Gamma(-2H,\overline{\tau})e^{\overline{\tau}}+\frac{\tilde{x}_0}{\overline{\tau}^{2H}}.\label{ExpressChi}
\end{align}
where $\Gamma(s,x)=\int_x^\infty t^{s-1}e^{-t}dt $ is the upper incomplete gamma function. We note that, the above expression satisfies the matching condition Eq.~(\ref{BehaviorChi1SmallArg}), suggesting that our analysis is consistent.  We also note that, when $t\to\infty$, the predictions of Eqs.~(\ref{ThreeTimes}) and (\ref{ExpressChi}) coincide  with the behavior (\ref{LongTimemPi}). This means that the complete structure of $m_\pi(t)$ has been determined, at all time scales. 

To evaluate the mean FPT, we introduce two intermediate time scales   $\varepsilon$, $\lambda$ that satisfy 
\begin{align}
t^*\ll\varepsilon\ll \tau_d \ll \lambda \ll T_{\text{RE}}. 
\end{align}
The mean FPT is evaluated by splitting the integral (\ref{MFPTx0}) over the three intervals $]0,\varepsilon[$, $]\varepsilon,\lambda[$ and $ ]\lambda,\infty[$, and by using the appropriate form of $m_\pi$ in Eq.~(\ref{ThreeTimes}) for each interval. This leads to
\begin{align}
\langle T\rangle p_s(L)=\frac{(t^*)^{1-H}}{\sqrt{\kappa}} \int_0^{\varepsilon/t^*} du \frac{e^{-f^2(u)/2u^{2H}}}{[2\pi u^{2H}]^{1/2}}  + \int_{\varepsilon}^{\lambda}  dt \left\{ \frac{e^{-[L(1-\phi)]^2/2\psi(t)}}{[2\pi\psi(t)]^{1/2}}-\frac{e^{-(L-x_0\phi(t))^2/2\psi(t)}}{[2\pi\psi(t)]^{1/2}}\right\} \nonumber \\
+\frac{A L T_{\text{RE}}^{1-2H}}{l^{3} \sqrt{2\pi}}  e^{-L^2/2l^2} \int_{\lambda/T_{\text{RE}}}^{\infty} du   \ L \left[\chi(u)- \frac{\tilde{x}_0}{u^{2H}}\right]  ,
\end{align} 
where for $t>\lambda$ we have used the fact that $m_\pi(t)\ll L$ and $x_0\phi\ll L$, and we have set $t=uT_{\text{RE}}$. Replacing $\chi$ by its value, and taking the limit $\lambda/T_{\text{RE}}\to0$ and $\varepsilon/t^*\to\infty$, we finally obtain 
\begin{align}
\langle T\rangle \simeq T_{RE} + T_{RE} ^{1-2H} \times \frac{A L(L-x_0)\Gamma(1-2H) }{l^2}, 
\end{align}
which is Eq.~(15) in the main text. 
  
 \subsection{Details on simulations and additional numerical controls}  
\label{FurtherChecks}
  
\begin{figure}[h!]
	\includegraphics[width=10cm]{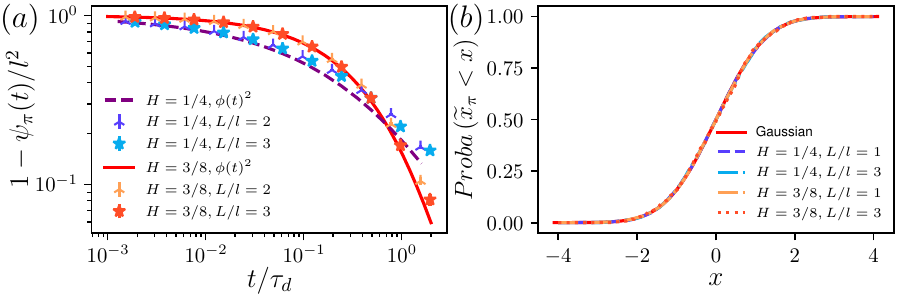}
	\caption{ (a) Additional check of the stationary covariance hypothesis. Here, $\psi_\pi(t)=\text{var}(x_\pi(t))$ and one represents $1-\psi_\pi(t)/l^2$ to determine whether the stationary covariance approximation is valid at long times (where $\psi_\pi(t)\to l^2$). Symbols are simulation results (parameter values are indicated in legend) and are compared to $\phi^2(t)$ obtained in the stationary covariance approximation (dashed and full lines). (b) Check of the Gaussian approximation. Here, one represents the cumulative distribution function (CDF) of the rescaled variable $\tilde{x}_\pi(t)=[x_\pi(t)-\langle x_\pi(t)\rangle]/\psi_\pi^{1/2}(t)$. The red line is the CDF of a normalized Gaussian. Other dashed lines represent simulation results, with parameters indicated in the legend. The collapse of the curves suggests that the stochastic process $x_\pi(t)$ is well approximated by a Gaussian process. }
	\label{FigCheck1}
\end{figure}

\begin{figure}[h!]
	\includegraphics[width=10cm]{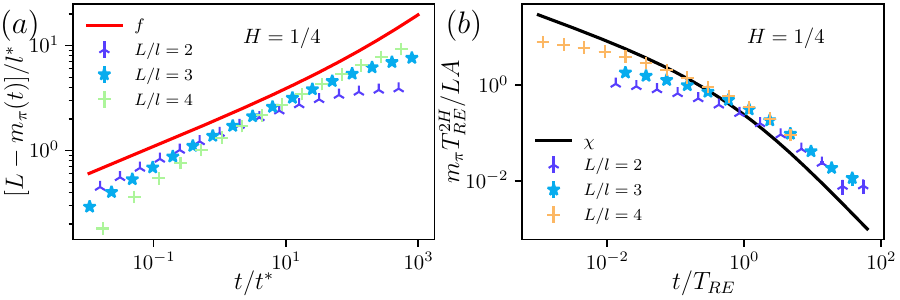}
	\caption{Additional checks for scaling behavior of $m_\pi(t)$ for $H=1/4$. (a) Check of the short time scaling (\ref{BehaviormPiShortTimes}) $m_\pi(t)=L-l^*f(t/t^*)$  in the limit $L\to\infty$. Here $f$ is calculated by numerically solving (\ref{Eq_for_f}). Note that the larger discrepancy between $f$ and the data at short times comes from the finiteness of the time step $\Delta t$ compared to $t^*$ (since $t^*\propto 1/L^4$ here). The fact that one needs to generate trajectories that are longer than $\langle T\rangle \propto e^{L^2/2l^2}$ prevents us from using smaller time steps for large $L$. (b) Check of the long time scaling regime given by Eq.~(13) in the main text. Here the initial position is drawn from an equilibrium distribution, corresponding to our predictions for $x_0=0$. }
	\label{FigCheck2}
\end{figure}

Here, we present additional numerical results supporting our findings. In Fig.~\ref{FigCheck1} we present additional tests of the validity of the Gaussian approximation and of the stationary covariance approximation. In Fig.~\ref{FigCheck2} we present a test of the scaling behavior of $m_\pi$ for large $L$. 
Last, we report the used values of the time step $dt$ for all simulations of this work in table \ref{tableTimeSteps}.

\begin{table}[h!]
\begin{center}
\begin{tabular}{| c | c | c |    c |}
 \hline
 $H$ & $L/l$ & $dt/\tau_d$ & Figures \\ 
 \hline
3/8 & 0 & $5.96\times 10^{-6}$ & Fig 2(a) \\  
3/8 & 1 & $5.96\times 10^{-6}$  & Figs. 2(a),  3(a), 4(b), 4(d), S1(b) \\  
3/8 & 2 & $5.96\times 10^{-6}$  & Figs. 2(a),  3(a), 4(a), 4(c), 4(d), S1(a)  \\  
3/8 & 3 & $7.45\times10^{-6}$ & Figs. 2(a),    3(a), 4(c), 4(d), S1(b)\\
3/8 & 3 & $1.49\times10^{-5}$ & Figs. 4(a), S1(a)\\  
3/8 & 4 & $1.86\times10^{-5}$  & Figs. 3(a), 4(d)\\  
3/8 & 4 & $7.45\times10^{-6}$  & 4(c)\\  
 \hline
1/4& 0 & $1.86\times 10^{-7}$ & Figs. 2(b) \\  
1/4& 1 & $1.86\times 10^{-7}$ & Figs. 2(b), 3(b), S1 \\  
1/4& 2 & $7.45\times 10^{-7}$ & Figs. 2(b), 3(b), 4(a), 4(b), S1, S2 \\  
1/4& 3 & $3.72\times 10^{-6}$ & Figs. 2(b), 3(b), 4(a), S1, S2 \\  
1/4& 4 & $1.30\times 10^{-5}$ & Figs. 3(b), S2 \\  
  \hline
\end{tabular}
\caption{Value of the time steps used in the simulations.}
\label{tableTimeSteps}
\end{center}
\end{table}

\subsection{Exactness of the theory at first order for weakly non-Markovian processes}
\label{Perturbation}

Let us consider the case of weakly non-Markovian processes, for which the covariance and mean of the process $x(t)$ are given by Eqs.~(\ref{IniMean}) and (\ref{IniCov}), with
\begin{align}
\phi(t)=e^{-\lambda t} + \varepsilon \phi_1(t),
\end{align}
with $\lambda>0$, $\varepsilon$ is a small parameter, and $\phi_1(t)$ is an arbitrary function. For simplicity, and without loss of generality, we set $\lambda=1$ and $l=1$.  We start with the generalization of Eq.~(\ref{Def2PointsPPi}) for an arbitrary number of positions and times $x_i,t_i$:
\begin{align}
p_\pi(L,t;&x_1,t+t_1;x_2,t+t_2;...;x_N,t+t_N)=\nonumber\\
&\int_0^\infty d\tau F(\tau)p(L,t+\tau;x_1,t+t_1+\tau;x_2,t+t_2+\tau;...;x_N,t+t_N+\tau \vert \text{FPT}=\tau) \label{DefNPointsPPi}.
\end{align}
Following the approach of Section \ref{DerivationEquations}, this equation leads to 
\begin{align}
\int_0^\infty dt [p_\pi(0,t; x_1,t+t_1; x_2,t+t_2;...)-p(0,t;x_1,t+t_1; x_2,t+t_2;...)] =    \langle T\rangle p_s(0;x_1,t_1; x_2,t_2;... ). 
\end{align}
We may write formally a continuous version of this equation, for all paths $[y(\tau)]$ with $y(0)=L$:
\begin{equation}
	\langle T\rangle P_{\text{s}}([y(\tau)])-\int_0^{\infty}dt \left\{\Pi([y(\tau)],t)-P([y(\tau)],t) \right\} = 0\label{Eq989},
\end{equation}
where $P_{\text{s}}([y(\tau)])$ is the stationary probability to follow the path $[y(\tau)]$, $\Pi([y(\tau)],t)$ is the probability to follow the path $[y]$ in the future $t$ of the FPT (ie, the probability density that $x(\text{FPT}+\tau+t)=y(\tau)$ for all $\tau>0$), and $P([y(\tau)],t)$ is the probability density that $x(t+\tau)=y(\tau)$ for all $\tau>0$. 
Using Bayes' formula, we can write   (\ref{Eq989}) as
\begin{equation}
	\int_0^{\infty}dt \left\{\Pi([y(\tau)],t\vert y(0)=L)p_\pi(L,t)-P([y(\tau)],t\vert y(0)=L)p(L,t) \right\} -\langle T\rangle p_s(L) P_{\text{s}}([y(\tau)]\vert y(0)=L)= 0,
\end{equation}
which is valid for all paths $[y(\tau)]$ (if $y(0)\neq L$ the above equation is simply $0=0$). Let us define a functional $\mathcal{F}([k])$ as the value of the above expression when multiplied by  $e^{\int_0^\infty d\tau k(\tau)y(\tau)}$ and integrated over all paths $y$. In principle $	\mathcal{F}([k])$ should vanish for all functions $k(\tau)$. Let us   evaluate $\mathcal{F}$ for a distribution of paths $\Pi$ that satisfies our hypotheses, i.e. by assuming that the process in the future of the first passage time is Gaussian with mean $m_\pi(t)$ and with the stationary covariance approximation. Using formulas for the moment generating function of Gaussian processes, we find
\begin{align}
&\mathcal{F}([k(\tau)])= \langle T\rangle p_s(L) e^{\int_0^\infty d\tau k(\tau)m_s^*(\tau)}e^{ \frac{1}{2}\int_0^\infty d\tau\int_0^\infty d\tau' k(\tau)k(\tau')\sigma(\tau,\tau')}\nonumber\\
&-\int_0^{\infty}dt  \left[p_\pi(0,t) e^{\int_0^\infty d\tau k(\tau) m_\pi^*(t+\tau\vert L,t)}-p(0,t) e^{\int_0^\infty d\tau k(\tau) m_0^*(t+\tau\vert L,t)}\right]e^{\int_0^\infty d\tau\int_0^\infty d\tau' \frac{k(\tau)k(\tau')}{2}\sigma(t+\tau,t+\tau'\vert t)},\label{ValueF}
\end{align}
where  we remind that $\langle T\rangle$ is evaluated with Eq.~(\ref{MFPTx0}), and
\begin{align}
\sigma(t+\tau,t+\tau'\vert t)=\sigma(t+\tau,t+\tau')-\frac{\sigma(t+\tau,t)\sigma(t+\tau',t)}{\sigma(t,t)}.
\end{align}
If one could find a function $m_\pi(t)$ so that $\mathcal{F}([k(\tau)])$ vanishes for all $k(\tau)$, it would mean that the theory is exact. It does not seem to be the case in general. However, when $\varepsilon\to0$, assuming that 
\begin{align}
m_\pi(t)=L e^{-\lambda t}+\varepsilon\mu_1(t),
\end{align}
we can evaluate (\ref{ValueF}) as 
\begin{align}
&\mathcal{F}([k(\tau)])=-\varepsilon \int_0^\infty d\tau k(\tau)Q_1(\tau) \times e^{\int_0^\infty du \int_0^\infty du'k(u)k(u')\frac{1}{2}\sigma(u,u')} +\mathcal{O}(\varepsilon^2),\label{DefinitionQ1Q2etc}
\end{align}
where
 \begin{align}
Q_1(\tau) = \int_0^{\infty}\frac{dt}{\sqrt{2\pi (1-e^{-2t})}}\Bigg\{ & e^{-\frac{[L(1-e^{-t})]^2}{2(1-e^{-2t})}}\left[\mu_1(t+\tau)-\mu_1(t)e^{-\tau}-Le^{-t}S_1(t,\tau)-L\phi_1(\tau)\right] \nonumber\\
&-e^{-\frac{(L-x_0e^{-t})^2}{2(1-e^{-2t})}}\left[m_1^*(t,\tau)-L\phi_1(\tau) \right]\Bigg\}  \label{EqMu1},
 \end{align}
where we have defined $S_1$ and $m_1^*$ such that
\begin{align}
&\frac{\sigma(t+\tau,t)}{\sigma(t,t)}=e^{-\tau}+\varepsilon S_1(t,\tau)+\mathcal{O}(\varepsilon^2), & m_0^*(t+\tau\vert L,t)=Le^{-\tau}+\varepsilon m_1^*(t,\tau)+\mathcal{O}(\varepsilon^2).
\end{align}
Note that, to obtain (\ref{DefinitionQ1Q2etc}), it is important to remark that 
\begin{align}
 \sigma(t+\tau,t+\tau'\vert t)=\sigma(\tau,\tau')+\mathcal{O}(\varepsilon^2).
\end{align}
We observe that the equality  $Q_1(\tau)=0$   for all $\tau$ can be realized by a proper choice of $\mu_1$, so that 
$\mathcal{F}([k(\tau)])$  \textit{vanishes at order $\varepsilon$ for all functions $k(\tau)$}. This suggests that  our theory is exact at order $\varepsilon$.

\end{document}